\documentclass[twocolumn,10pt,times,epsfig]{IEEEtran} 
\usepackage{epsfig,amsmath,url,bm}
\usepackage{epstopdf}
\usepackage{amsmath,amssymb,subfigure,graphicx}
\usepackage{amsthm}
\usepackage{caption}
\usepackage{xfrac}
\usepackage{utopia}
\usepackage{multirow}
\usepackage[backend=bibtex,sorting=none]{biblatex}
\begin{document}
\title{Efficient Detection Of Infected Individuals using Two Stage Testing}
\author{Arjun Kodialam \\ Marlboro High School \\ Marlboro, NJ 07746}
\date{}
\maketitle

\section{Abstract}
Group testing is an efficient method for testing a large population to detect infected individuals. In this paper, we consider an efficient two stage group testing scheme.
Using a straightforward analysis, we characterize the efficiency of several two stage group testing
algorithms. We determine how to pick the parameters of the tests optimally for three schemes with different types of randomization, and show that the performance of two stage testing depends on the type of randomization employed.  Seemingly similar randomization procedures lead to different expected number of tests to detect all infected individuals, we determine what kinds of randomization are necessary to achieve optimal performance. We further show that in the optimal setting, our testing scheme is robust to errors in the input parameters.

\section{Introduction}
There has been a huge surge of interest in group testing to facilitate testing a large number of patients for COVID-19 while testing supplies are limited.  Several countries like Germany have begun using group testing techniques to reduce the total number of tests needed to identify infected individuals \cite{Bui2020Jul}. Similar algorithms have also been proposed as a mechanism to test populations at public facilities like airports and schools.

Group testing \cite{dorfman1943detection} was first utilized in WWII to detect diseases among soldiers. The method is based upon the creation of pooled samples that are mixtures of samples from multiple individuals.
If a disease test run on the pooled sample is negative, then we conclude that none of the individuals in the group have the disease. If the test is positive, then we know that at least one of the individuals in the group is infected, and thus must be tested further. There are many different ways of performing group testing \cite{Mallapaty2020Jul} and each method has its own advantages and drawbacks -- two commonly used metrics to evaluate the efficiency of group testing procedures are 
\begin{itemize}
    \item The total number of tests needed to establish which members of the population have a disease
    \item The amount of time needed to perform the tests
\end{itemize} 
The amount of time required to perform the tests depends on the number of tests that can be done in parallel. We call each set of parallel tests a \emph{stage}. In single stage testing, all tests are done in parallel and at the end of these tests the infected individuals are identified. 

A simple example of single stage testing is when a sample from each individual is separately tested in parallel. The results of this set of tests will clearly identify which individuals are infected. However, if all infected individuals have to be identified with certainty a single stage testing algorithm would use a number of tests equal to the total number of people in the population of interest. This makes single stage testing very expensive, especially when the population is large and the number of infected individuals is a small fraction of the population. 

In this paper, we consider two stage testing schemes, where at the cost of taking an extra round of testing, the number of tests is reduced significantly compared to single stage testing. Theoretical analysis finds that from a population of $n$ individuals out of which $k$ are infected, two stage testing requires $O \left( k \log \left( \frac{n}{k} \right) \right)$ tests (as opposed to $n$ tests for single stage testing) \cite{knill1995lower}. While prior research has considered cases in which we merely wish to determine which people are \emph{likely} to be infected with some probabilistic threshold, in this paper we assume that all the infected individuals have to be identified with certainty. We further assume that our tests are reliable enough that false positive and negative results are negligible. 

\section{Model and Objective}
We consider a population of $n$ individuals, out of which $k \ll n$ are infected. The value of $n$ and $k$ are assumed to be known to the test designer. The results derived in this paper also carry over to the binomial infection model where instead of $k$ being fixed, each individual is assumed to be infected with probability $p$. 

The objective is to identify {\em all} the infected individuals using two stage testing. 
Since our testing scheme is inherently probabilistic, the number of tests needed to identify the infected individuals will be a random variable $T$ and our objective is to minimize the expectation of the total number of tests needed in the first and second stage combined. 

\subsection{Related Work}
Group testing can be broadly classified into adaptive and non-adaptive testing.
In adaptive testing, the test pools are designed sequentially, and the 
design of a test pool can depend on the result of the previous tests. In non-adaptive testing, all the test pools are designed in advance and therefore tests can be done in parallel. This leads to a significant reduction in testing time, but reduces efficiency in terns of number of tests by limiting the test designer's ability to use information dynamically. 
There is large body of work in non-adaptive group testing \cite{Du1999Dec} that has concentrated on carefully
constructing test designs with the property that that the test can deterministically determine the infected individuals if the number of infected individuals is known with certainty or can be upper bounded. 
However, such designs are generally not practical due to the fact that if the bounds are violated even slightly, the entire test can fail. Moreover, the number of tests required for these test designs is significantly more than the expected number of tests that we require for probabilistic methods. 

 
 The design of pooled tests depends on the optimization objective and the infection model. We consider two such models. First, in the \emph{simple infection model}, we assume that the number $k$ of infected individuals is given. The probabilistic methods developed in this paper are robust to errors in the estimation of the number of infected individuals. The second model is the \emph{binomial infection model}, where it is assumed that each member of the population is infected with probability $p$. The size of the population $n$ and the number $k$ or infection probability $p$ are known to the test designer. If the fraction of infected individuals is high, then two stage group testing does not give any benefit compared to exhaustive testing. In fact \cite{Fischer1999Jan} shows that if the fraction of infected individuals is greater than $\frac{3 - \sqrt{5}}{2} \approx 0.382$, then the expected number of tests using two-stage testing is greater than $n$.
 
 Lower bounds on the expected number of tests for two stage testing were derived in \cite{knill1995lower}, and using information theory techniques \cite{aldridge2019group} asymptotic bounds on the expected number of tests under the binomial model can be computed  \cite{berger2002asymptotic} by assuming that the probability of infection $p=n^{-\beta}$ for $\beta \geq 0$, and considering the limit $n \rightarrow \infty$.  This result  was improved using statistical physics techniques in \cite{mezard2007group} where it was shown that the asymptotic efficiency is lower bounded by $\frac{1}{\log^2 2}$ if $0 \leq \beta < \frac{1}{2}$. For  $\beta \geq \frac{1}{2}$ the asymptotic efficiency is between $\frac{1}{\log^2 2}$ and $e$. As in \cite{mezard2007group} we measure the asymptotic efficiency of our testing algorithms by the ratio of the expected number of tests to $k \log{\left(\frac{n}{k} \right)}.$ 
 
 Two stage testing models like the ones considered in this paper are also considered in \cite{aldridge2020conservative}.  The paper does not consider the optimization of the parameters of the test allocation mechanism. Therefore, they do not achieve the efficiency that we achieve in this paper. In fact, 
  the expected number of tests needed in our fixed tests per individual (FTI) two stage pool design outlined in Section \ref{fti} is $\frac{1}{\log^2 2} \;  k \; \log{\left(\frac{n}{k} \right)} + O(k)$ suggesting that it is efficient based on the lower bound computed in \cite{mezard2007group}.
 
 \subsection{Our Contribution}
 The main objective of this work, is to  develop a two stage group testing protocol that is simple to implement and reduces the number of tests while keeping the time to detect the infected individuals low. In addition, the result of the testing procedure is easy to interpret and implement in real settings like schools.  The paper makes the following contributions:
 \begin{itemize}
     \item A simple analysis of the various randomization schemes for two stage group testing that permits the optimization of the design parameters.
     \item A justification of why seeming similar randomization schemes lead to different expected number of tests. For instance, we show that performing a constant number of tests per individual performs better in theory and practice than a constant number of tests per pool.
     \item We outline how the results that are derived for a known value of $k$ is very robust to errors in the estimation of $k$. In fact, we show that the results in this paper carry over directly to the the binomial infection model where the number of infected individuals $k$ is replaced with the expected number of infected individuals ${\overline{k}} = np$.
     \item We show an excellent match between theoretical results and simulations for all results in the paper.
 \end{itemize}

\section{Two Stage Testing}
\label{twostage}
In our two stage testing schemes, the first stage is a screening stage that identifies a set of potentially infected individuals. The efficiency of the two state testing protocol relies on being able to detect potentially infected individuals with as few tests as possible. 
Overall, the two stage testing scheme works as follows:
\begin{itemize}
    \item {\bf Screening Stage:} In the first stage, a set of pooled tests are conducted in parallel. The pooled tests will result the identification of a set of potentially infected people. 
    \item{\bf Confirmation Stage:}  In the second stage, all potentially infected people as determined in the first stage are tested individually in parallel, and the truly infected people are identified in this testing cycle.
\end{itemize}
Recall that in a pooled testing \cite{dorfman1943detection} scheme, samples from multiple individuals are 
mixed together and a single test is performed on the mixture, and that a positive result will be indicated by the test if {\em any one} of the individuals in the mixture has the virus. The test will indicate a negative if {\em all} individuals in the mixture are uninfected. As in all pooled tests, we assume that the dilution due to pooling effects do not affect the test results -- for the specific use-case of COVID-19, this assumption is supported by the methods of researchers in Germany, who pool as many as $30$ samples at a time \cite{Bui2020Jul}. In further research, we will design pooled tests that take into consideration the testing inaccuracy that could result from sample dilution.
The main decisions that have to be made when performing pooled testing are:
\begin{itemize}
    \item The number of pooled tests $m$ done in the first stage.
    \item The pooling scheme, that determines which pool or pools to which an individual contributes his samples in the first stage.
\end{itemize}
\subsection{Determining Potentially Infected Individuals}
In the first stage,  an individual is generally part of multiple pools. If {\bf any} of the pools that the individual is part of is negative, then the 
individual is not infected and is cleared. If {\bf all} the tests that a person is part of is positive, then that person is potentially infected.
These people will be tested individually in the second phase. Note that all the pools that an infected individual is part of will test positive and all these infected individuals will be tested in the second phase. We now consider three different pooling schemes in the next three sections. 
\section{Fixed Number of Tests Per Test Pool (FTP)}
\label{ftp}
We consider a two stage testing scheme where we have $m$ first stage pools and each pool picks $b$ individuals out of $n$ at random. We call this scheme fixed number of tests per pool (FTP). 
\subsection{Illustrative Example}
\begin{figure}[h]
\centering
\includegraphics[width=2.5 in]{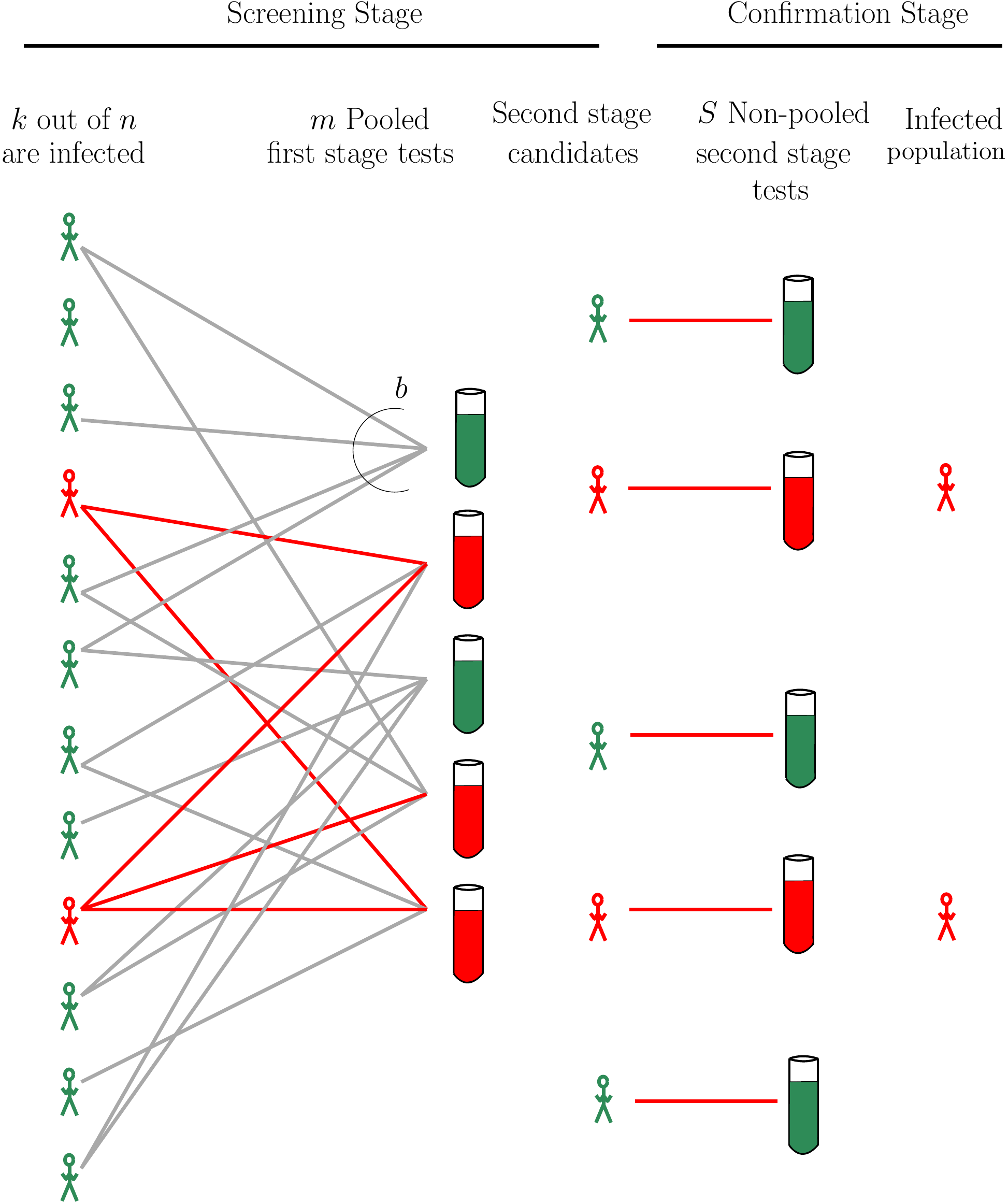}
\vspace{.2mm}
\caption{Fixed Number Tests Per Pool (FTP) with $m=5$ pools and $b=4$ tests per pool.}
\label{gtest6}
\end{figure}

We illustrate the FTP pooling scheme in Figure \ref{gtest6}. In this example, there are 
$12$ individuals in the population of which $2$ (Numbers $4$ and $9$) are infected (shown in red). 
There are $m=5$ pooled tests in the first stage and each pool picks  $b=4$ people randomly. The $3$ pools that will show a positive result are shown in red. There will be a positive result if any infected individual 
is in the pool. The first stage picks the infected individual as well as non-infected individuals (Number $2,7$ and $11$ in the population) for whom all tests are positive. Note that $2$ is not tested in the first stage due to the randomness of the choice of the individuals. These $5$ individuals are tested in the second stage to determine the 
infected individuals. Note that this scheme has a total of $10$ tests ($5$ in the first stage and $5$ in the second stage).  For large $n$ and small $k$, a carefully designed two stage scheme will result in a significant reduction in the number of tests needed to identify the infected individuals.  

Fix an individual, and we want to compute the probability that this individual is not picked by a pool.
Since, each pool picks $b$ out of $n$ individuals the probability $q$ that a particular individual is not picked by a given 
pool is 
\begin{equation}
\label{zbnd}
q= \frac{{n-1 \choose b}}{{n \choose b}} =\left( 1- \frac{b}{n} \right).
\end{equation}
The denominator of the first term is the number of different ways of choosing $b$ individuals out of $n$. 
Since we do not want any pool to pick this specific individual, the numerator is the number of different ways of picking $b$ individuals out of the remaining $n-1$ individuals.
There are $k$ infected individuals in the population. Each pool picks its participants independently of all other pools.  The probability that a particular pool 
tests negative $\omega_n$ in the first phase equals the probability that that pool does not pick any infected 
individuals. 
\begin{equation}
\omega_n = \left( 1 - \frac{b}{n} \right)^k \approx e^{\frac{-kb}{n}}
\label{omega}
\end{equation}
(where we use the approximation $1-x \approx e^{-x}$ when $x$ is small) and $\omega_p = 1- \omega_n.$ 
The probability that an individual tests positive given that $j$ pools pick the individual is $\omega_p^j.$ The probability that 
$j$ pools pick an individual is 
${m \choose j} \left(\frac{b}{n} \right)^j \left(1-\frac{b}{n} \right)^{n-j}$. Therefore, the probability that an individual tests 
positive is 
\begin{eqnarray*}
t_p & \sim & \sum_{j=0}^{m} { m \choose j} \left(\frac{b}{n} \right)^j \left(1-\frac{b}{n} \right)^{n-j} \omega_p^j \\
& = & \left( \left( \frac{b}{n} \right) \omega_p + \left( 1-\frac{b}{n} \right) \right)^m \\
& = & \left(\left( \frac{b}{n} \right)  \left(1-e^{-kb/n} \right) + \left(1-\frac{b}{n} \right) \right)^m \\
& = & \left( 1 -\frac{b}{n} \; e^{-kb/n} \right)^m
\end{eqnarray*}
The reason that the first line is an approximation is that we assume that the probability that $j$ pools are positive is $\omega_p^j$. This assumes that the pools are independent which they are not. However, for a large number of pools the dependence is quite weak.  Therefore the approximation is quite accurate.
The expected number total number of tests is 
\begin{equation}
E[T]= m +  k + (n-k) \left( 1 -\frac{b}{n} \; e^{-kb/n} \right)^m.
\label{expt}
\end{equation}
We want to choose $m$ and $b$ to minimize $E[T]$
\subsection{Choosing $m$ and $b$ to Minimize the Expected Number of Tests}
We first assume that the value of $m$ is fixed and find the value of $b$ that minimizes the expected number of tests as a function of $m$. We define
$$G(b)  =   \left( 1 -\frac{b}{n} \; e^{-kb/n} \right).$$
Then $E[T]= m + k + (n-k) G(b)^m.$ For a fixed $m$, we want to find $b^*$ that minimizes $G(b).$ 
We compute the derivative of $G(b)$ with respect to $b$ and set it to zero and solve for $b$.
$$\frac{\partial G(b)}{\partial b} = \left( \frac{b}{n} \right) \left( \frac{k}{n} \right)
e^{-kb/n} - \left( \frac{1}{n} \right) e^{-kb/n} =0.$$
Therefore 
$$b^* = \frac{n}{k}.$$
It is easy to check that this is the minima.
Substituting $b^*$ in the expression for $G(b)$, we get 
\begin{equation}
E[T] = m + k + (n-k) \; \left(1 - \frac{1}{ek} \right)^m.
\label{expt3}
\end{equation}
We want to determine $m$ that minimizes the above expression.
Differentiating Equation (\ref{expt3}) with respect to $m$ and setting to zero, 
we get
\begin{equation}
\frac{\partial E[T])}{\partial m} = 1 + (n-k)\left(1 - \frac{1}{ek} \right)^m 
\log \left(1 - \frac{1}{ek} \right)=0.
\label{diff}
\end{equation}
Solving for the optimal value of $m$, we get 
\begin{eqnarray*}
m^* & = &  \frac{ \log \left[ \frac{-1}{(n-k) \log \left(1 - \frac{1}{ek} \right)} \right] }{\log \left(1 - \frac{1}{ek} \right)} \\ 
& \approx &  e k \log \left( \frac{n-k}{e k} \right)\\
& = & e k \log \left( \frac{n-k}{k} \right) - e k
\end{eqnarray*}
where we have approximated
\begin{equation}
\log \left(1 - \frac{1}{ek} \right) \approx \frac{-1}{ek}. 
\label{approx1}
\end{equation}
From Equation (\ref{diff}) note that
$$\left(1 - \frac{1}{ek} \right)^m = \frac{-1}{(n-k) \log \left(1 - \frac{1}{ek} \right)}.$$
Substituting this into $E[T]$ we get 
\begin{eqnarray}
E[T] & = & m + k + (n-k) \; \left(1 - \frac{1}{ek} \right)^m  \nonumber \\
& \approx & e k \log \left( \frac{n-k}{k} \right) - e k + k + \frac{-1}{\log\left( \left(1 - \frac{1}{ek} \right) \right)}  \nonumber \\
& \approx & e k \log \left( \frac{n-k}{k} \right) - e k + k + \frac{-1}{- \frac{1}{ek} }  \nonumber \\
& \approx & k + e \;  k \log \left( \frac{n-k}{k} \right) 
\label{expval1}
\end{eqnarray}
where we again use the approximation in Equation (\ref{approx1}).
In Section \ref{simul}, we show the excellent agreement between the 
theoretical expected value in Equation (\ref{expval1}) and simulation
results. 
We now outline a second pooling scheme where instead of each
pool choosing $b$ individuals at random, each individual picks $d$ tests at random.

\section{Fixed Number of Tests Per Individual (FTI)}
\label{fti}
Instead of a fixed number of 
tests per pool, we now consider a pooling design where each individual picks a fixed number of pools to participate in.
Assume that there $m$ pooled samples in the first stage of the testing protocol and each individual picks $d$ pools at random from the $m$ pools.
We call this protocol fixed number of tests per individual (FTI). We now illustrate FTI with an example.
\subsection{Illustrative Example}
\begin{figure}[h]
\centering
\includegraphics[width=2.5 in]{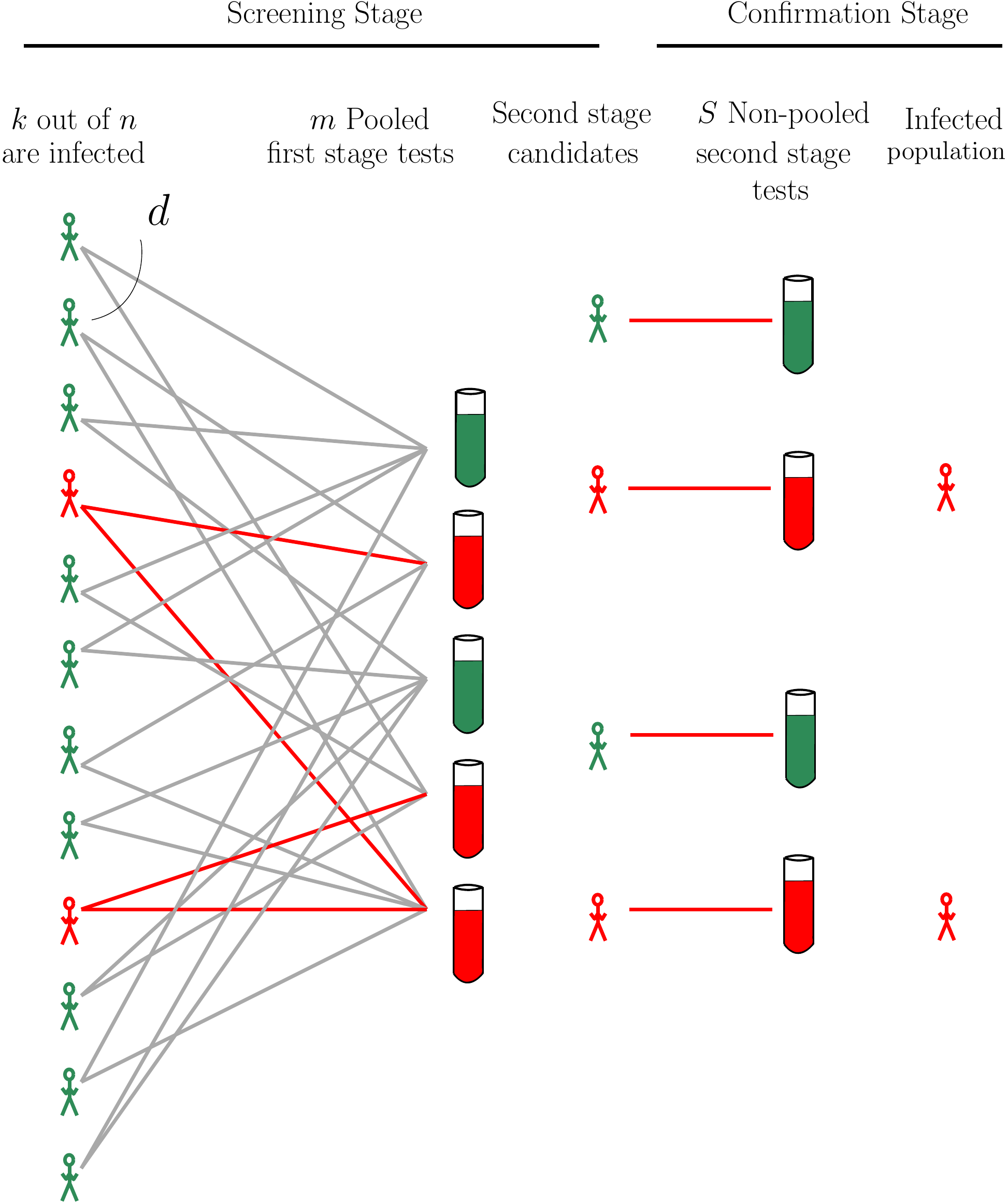}
\vspace{.2mm}
\caption{Fixed Number Tests Per Individual (FTI) with $m=5$ pools and $d=2$ tests per individual.}
\label{gtest}
\end{figure}

We illustrate two stage testing in Figure \ref{gtest}. In this example there are 
$12$ individuals in the population of which $3$ (Numbers $4$ and $9$ in the population) are infected (shown in red). 
There are $m=5$ pooled testing in the first stage and each individual participates in $d=2$ pools picked at random. The $3$ pools that will show a positive result are shown in red. There will be a positive result if any infected individual 
is in the pool. The first stage picks the infected individual as well as non-infected individuals (Number $2$ and $7$ in the population) for whom both tests are positive. These $4$ individuals are tested in the second stage to determine the 
infected individuals. Note that this scheme has a total of $9$ tests ($5$ in the first stage and $4$ in the second stage).  For large $n$ and small $k$, a carefully designed two stage scheme will result in a significant reduction in the number of tests needed to identify the infected individuals.  
\subsection{Total Number of Tests}
We assumed that there are $m$ pooled tests in the first stage and this identifies 
a set of potentially infected individuals $S$ who will be tested in the second 
stage. Therefore the total number of tests $T=m+S$.
We want to determine a $(m,d)$ scheme that minimizes the expected number of tests
$E[T]= m + E[S]$.
\subsection{Minimizing the Expected Number of Tests}
Each individual picks $d$ out of $m$ pools at random in the first phase.
Therefore, the probability $q$ that a particular pool is not picked by a given 
individual is 
\begin{equation}
\label{zbnd2}
q= \frac{{m-1 \choose d}}{{m \choose d}} =\left( 1- \frac{d}{m} \right).
\end{equation}
The denominator of the first term is the number of different ways of choosing $d$ pools out of $m$. 
Since we do not want any and individual to pick a specific pool, the numerator is the number of different ways of picking $d$ pools out of the remaining $m-1$ pools.
There are $k$ infected individuals in the population. Each individual picks her set of pools independently of all other individuals.  The probability that a particular pool 
tests negative in the first phase equals the probability that that pool is not 
picked by any of the infected individuals. Each individual does not pick this pool 
with probability $q$. The probability that none of the $k$ infected 
individual picks this pool is given by $q^k$  since the picks are independent. 
Therefore the probability $\phi_n$ that 
a particular pool is negative is 
$$\phi_n = \left( 1- \frac{d}{m} \right)^k \approx e^{-kd/m}$$
where we used the standard approximation that $e^x \approx 1-x$ if $x$ is small. Since the value of $m$ will be large compared to $d$, the approximation holds true for the most common cases.
The probability that a particular pool is positive $\phi_p = 1 - \phi_n.$
The probability that a particular individual is identified as potentially positive in the first phase is when all the $d$ pools picked by the individual are positive. In the case of an infected individual (there are $k$ of them) all the pools that they belong to will be infected. For the other
$n-k$ non-infected individuals, the probability that they will  be identified as potentially infected is the probability that all the pools picked by them
are positive. This probability that a pool tests positive is  $\phi_p$ for a given pool and the probability that all $d$ pools test positive is $\phi_p^d$. This is an approximation since the information that one pool is positive changes the probability that some other pool is positive, but for a large number of pools the dependence is weak.
Let $S$ denote the (random) set of individuals selected as potentially positive in the first stage.
Then,
$$E[S] \sim  k + (n-k) \phi_p^d = k + (n-k) \left( 1 - e^{-kd/m} \right) ^d.$$
Let $T$ denote the total number of first and second stage tests. The expected number total number of tests is 
\begin{equation}
\label{exptest}
E[T]= m +  k + (n-k) \left( 1 - e^{-kd/m} \right)^d.
\end{equation}
We want to choose $m$ and $d$ to minimize $E[T]$
\subsection{Choosing $m$ and $d$ to Minimize the Expected Number of Tests}
We first assume that the value of $m$ is fixed and find the value of $d$ that minimizes the expected number of tests as a function of $m$. We define
$$f(d)  =  \left(  1- e^{-\frac{kd}{m}}  \right)^d.$$
Then $E[T]= m + k + (n-k) f(d).$ For a fixed $m$, we want to find $d^*$ that minimizes $f(d).$ We rewrite
$$f(d) =  e^{ - \frac{m}{k} \log \left[ \left(  1- e^{-\frac{kd}{m}} \right) e^{-\frac{kd}{m}} \right] } $$
 This expression is minimized when 
$$ \log \left[ \left(  1- e^{-\frac{kd}{m}} \right) e^{-\frac{kd}{m}} \right] $$
is maximized. This occurs when 
$$e^{-\frac{kd}{m}} = \frac{1}{2}$$
or when 
\begin{equation}
\label{defd}
d^*= \frac{m}{k} \log 2
\end{equation}
as shown in \cite{broder2004network}.
Therefore 
$$f(d^*) = \left[ \left( \frac{1}{2} \right)^ {\log 2} \right]^{\frac{m}{k}}.$$
For convenience, we set 
\begin{equation}
\label{defb}
\beta = \left( \frac{1}{2} \right)^ {\log 2}.
\end{equation}
Note that $ \log \beta = - \log^2 2.$
Substituting $\beta$  into Equation (\ref{exptest}), we get
\begin{eqnarray*}
E[T] & = & m + k + (n-k) \; \beta^{\frac{m}{k}} \\
& = & k \left[ \frac{m}{k} + 1 + \left( \frac{n-k}{k} \right) \;  \beta ^{\frac{m}{k}} \right].
\end{eqnarray*}
We want to determine $m$ that minimizes the above expression.
Setting $y= \frac{m}{k}$ and 
$$g(y) = 1+ y + \left( \frac{n-k}{k} \right) \; \beta^y,$$
note that $E[T]= k g(y).$ We want to determine $y^*$ that minimizes $g(y)$. Differentiating 
$g(y)$ with respect to $y$ and setting to zero gives
$$\frac{\partial g(y)}{\partial y} = 1 + \left( \frac{n-k}{k} \right) \; \beta^y \log \beta = 0.$$
Solving for $y$ to get
\begin{equation}
\label{defy}
y^* = - \log \left[ - \left( \frac{n-k}{k} \right) \log \beta \right] \frac{1}{\log \beta}
\end{equation}
and the optimal value of $m$ denoted by $m^* = k y^*$ 
$$
m^* = - k \log \left[ - \left( \frac{n-k}{k} \right) \log \beta \right] \frac{1}{\log \beta}
$$
Substituting $\log \beta = - \log^2 2$, we get 
\begin{eqnarray}
m^* & = & \frac{k}{\log^2 2} \log \left[ \left( \frac{n-k}{k} \right) \log^2 2 \right]  \nonumber \\
& \approx & 2.08 \; k \log \left[ 0.48 \left(  \frac{n-k}{k} \right)  \right] \label{defm}
\end{eqnarray}
Substituting the optimal value of $y^*$ to compute 
$g(y^*)$, we obtain
$$g(y^*) = 1 - \log \left[ - \left( \frac{n-k}{k} \right) \log \beta \right] \frac{1}{\log \beta} - \frac{1}{\log \beta}.$$
Plugging in $\log \beta = - \log^2 2$ we get
$$g(y^*) = 1 + \frac{1}{\log^2 2} \log \left[ \left( \frac{n-k}{k} \right) \log^2 2  \right]  + \frac{1}{\log^2 2}.$$
Therefore, the expected number of tests is 
\footnotesize
\begin{eqnarray*}
E[T] & = & k \left( 1 + \frac{1}{\log^2 2} \log \left[ \left( \frac{n-k}{k} \right) \log^2 2  \right]  + \frac{1}{\log^2 2} \right) \\
& = & \left( 1 + \frac{1}{\log^2 2} + \frac{2 \log \log 2}{\log^2 2} \right) k  + \frac{1}{\log^2 2} k \log \left( \frac{n-k}{k} \right)
\end{eqnarray*}
\normalsize
Note that the coefficient of the $\log \left( \frac{n-k}{k} \right)$ term  is 
$\frac{1}{\log^2 2}$ and this matches the lower bound on the asymptotic efficiency computed
in \cite{mezard2007statistical}. 
Computing the approximate numerical values, we get 
$$E[T] = 1.55 k + 2.08 k \;  \log \left( \frac{n-k}{k} \right).$$
It is easy to show that the expected number of tests required for FTI is less than the expected number of tests for FTP for all values of $n$ and $k$. 
Another commonly studied two stage group testing scheme is randomized Pooling which we study next.

\section{Randomized Pooling}
Like FTP and FTI schemes, randomized pooling has $m$ pools in the first stage. Each individual is placed into each pool with probability $0 \leq a \leq 1$. This scheme can be implemented in one of two ways: Each individual goes though the pools one at a time and picks each pool with probability $a$. Alternatively, each pool can go through each individual and pick each individual with probability $a$. In the description we assume the first implementation where the choice is made be the individual. As in the last two sections, the values of $m$ and $a$ will be chosen to minimize the expected number of tests. 
Each individual picks each pool with probability $a$. 
There are $k$ infected individuals. Since they pick each pool independently,
the probability that none of the $k$ infected
individual picks this pool is given by $(1-a)^k$.  
Therefore the probability $\theta_n$ that 
a particular pool is negative is 
$$\theta_n = \left( 1- a \right)^k \approx e^{-ak}$$
where we used the standard approximation that $e^x \approx 1-x$ if $x$ is small. We assume that $p$ is small and  the approximation holds true for the most common cases.
The probability that a particular pool is positive $\theta_p = 1 - \theta_n.$
The probability that a particular individual is identified as potentially positive in the first phase is when all the pools picked by the individual are positive or the individual does not participate in any pool. In the case of an infected individual (there are $k$ of them) all the pools that they belong to will be infected. For the other
$n-k$ non-infected individuals, the probability that they will  be identified as potentially infected is the probability that all the pools picked by them
are positive. 
We now compute the probability that an individual tests positive.
Assume that the individual picks $j$ tests. This occurs with probability 
$${ m \choose j} a^j (1-a)^{m-j} \quad j=0,1,2, \ldots m.$$
If she picks $j$ tests then the probability that she tests positive
is $\theta_p^j.$ Therefore the probability that an individual tests positive $t_p$ is given by 
\begin{eqnarray*}
t_p & \sim & \sum_{j=0}^{m} { m \choose j} a^j (1-a)^{m-1} \theta_p^j \\
& = & \left(a \theta_p + (1-a) \right)^m \\
& = & \left(a (1-e^{-ka}) + (1-a) \right)^m \\
& = & \left( 1 - a \; e^{-ka} \right)^m
\end{eqnarray*}
Let $T$ denote the expected number of tests. 
Then,
$$E[T]= k + (n-k) t_p = k + (n-k) \left( 1 - a \; e^{-ka} \right)^m.$$
As in the last two sections, we first fix $m$ and solve for $a$ as a function of $m$.
 We define
$$G(a)  =  \left( 1 - a \; e^{-ka} \right).$$
Then $E[T]= m + k + (n-k) G(a)^m.$ For a fixed $m$, we want to find $a^*$ that minimizes $G(a).$ 
We compute the derivative of $G(a)$ with respect to $a$ and set it to zero and solve for $a$.
$$
\frac{\partial G(a)}{\partial a} = ka e^{-ka} - e^{-ka} =0.$$
Therefore 
$$a^* = \frac{1}{k}.$$
It is easy to check that this is the minima.
Substituting this value into the expression for $E[T]$, we get
\begin{equation}
E[T] = m + k + (n-k) \; \left(1 - \frac{1}{ek} \right)^m.
\label{expt4}
\end{equation}
This equation is exactly the same as Equation (\ref{expt3}) and the rest of the derivation to obtain the optimum value of $m$ is the same at the FTP scheme. Therefore the expected number of tests for random pooling is 
exactly the same FTP. 

Picking a fixed number of tests per individual outperforms both fixed number of tests per pool as well as random pooling. So far, we have assumed that the value of $k$ is known to the test designer. We now show how all the results derived so far extend approximately to the binomial infection model. 
\section{Binomial Infection Model}
\label{binmodel}
In the binomial infection model, each of the $n$ individuals is assumed to be infected with probability $p$ and let $\overline {k}= np$ denote the expected 
number of infected individuals.
The results derived in the last two section can be extended approximately to the binomial infection model. For the FTP testing scheme, consider the computation of the probability that a pool does not pick any infected individuals shown in Equation (\ref{omega}). In the binomial infection model,
the number of infected individuals is a random variable. The probability that 
a pool does not pick any infected individuals is 
\begin{eqnarray*}
\omega_n & = & \sum_{j=0}^{n} { n \choose j} p^j (1-p)^{n-j} \left(1-\frac{b}{n} \right)^j \\
& = & \left( p \left( 1 - \frac{b}{n} \right) + 1 - p \right)^n \\
& = & \left( 1 -\frac{pb}{n} \right)^n \\
& \approx & e^{-pbn/n} = e^{-b\overline {k}/n}
\end{eqnarray*}
Therefore, in the expression for the probability that a pool does not pick an infected individual, we can replace the fixed number of infected individual $k$ with
the expected number of infected individuals $\overline {k}$. In the expression for the expected number of tests in Equation (\ref{expt}), there is a product of the number of uninfected individuals and the probability that an uninfected individual is identified as positive. 
These two random variables are not independent. However, the probability that a pool is positive is highly concentrated around its mean (see \cite{broder2004network} for a similar argument for the Bloom Filter). Therefore the probability that is positive can be taken as approximately independent of the actual number of infected but only depends on the mean number of infected individuals. In this case, we can replace $k$ with ${\overline{k}} = np$ and write Equation (\ref{exptest}) as
\begin{equation}
\label{exptest_bin}
E[T] \approx  m +  {\overline{k}} + (n-{\overline{k}}) \left( 1 - e^{-{\overline{k}}d/m} \right)^d.
\end{equation}
The rest of the derivation follows exactly the same steps as FTI and after optimizing the parameters, we get the expected number of tests for FTI in the binomial case as approximately 
\footnotesize
\begin{equation}
E[T]
 \approx \left( 1 + \frac{1}{\log^2 2} + \frac{2 \log \log 2}{\log^2 2} \right) np  + \frac{1}{\log^2 2} n p \log \left( \frac{1-p}{p} \right)
\end{equation}
\normalsize
where $k$ in Equation (\ref{exptest}) is replaced by $np$.
Using the same argument, the expected number of tests for FTD and Random Pooling for the binomial case 
$$E[T] \approx  np + e \;  np \log \left( \frac{1-p}{p} \right) $$

\section{Performance of the FTI and FTP Algorithms}
\label{simul}
Though all the randomization approaches yield $O\left( k \log \frac{n}{k} \right)$
expected number of tests, the number of tests required varies according to the type of randomization  used. In this section, we simulate the testing algorithms for two different population sizes $n=1000$  and $n=10000$ and vary the number of infected individuals. 
\subsection{Simulation with Fixed $k$}
\begin{figure}[h]
\centering
\includegraphics[width=2.5 in]{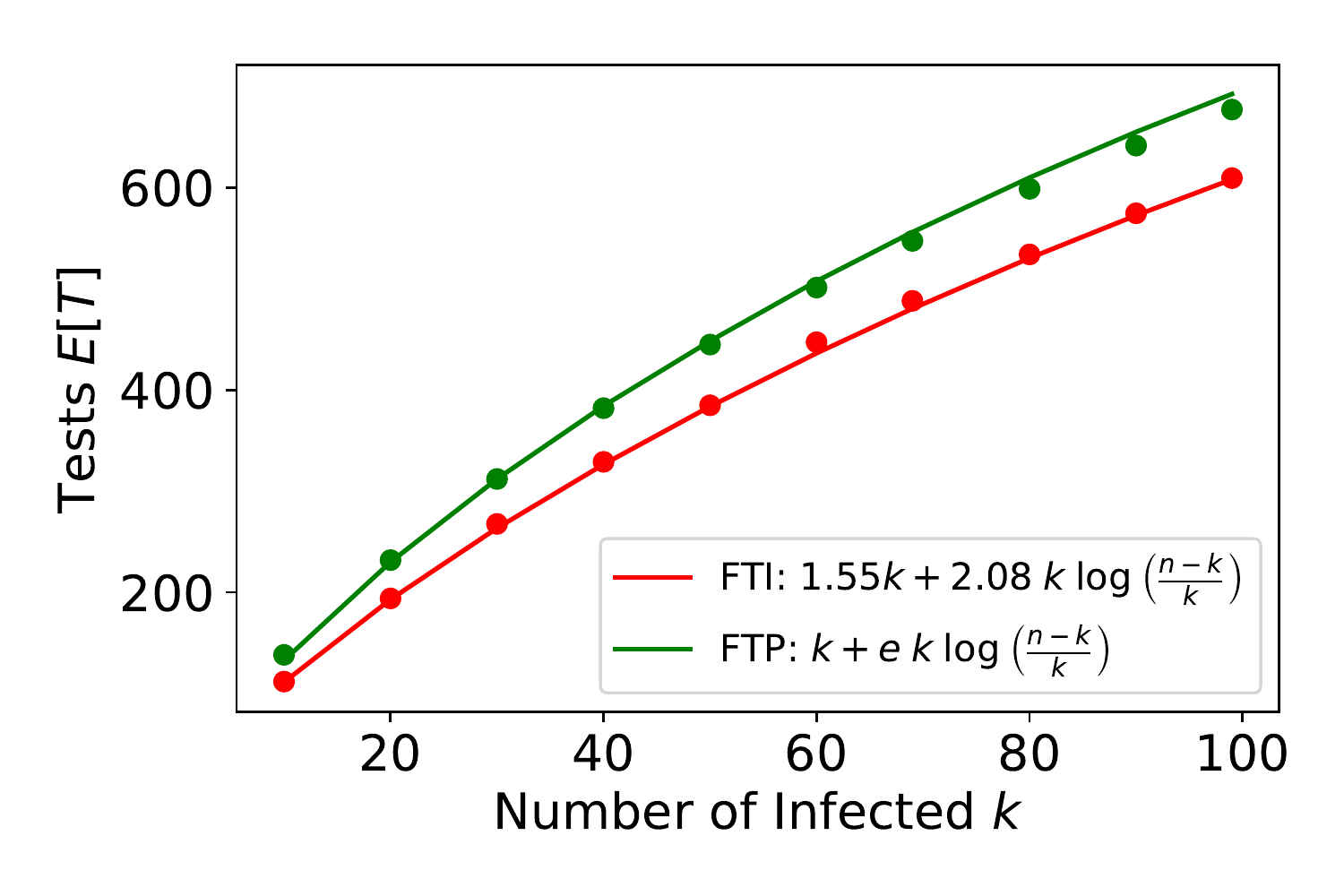}
\vspace{.2mm}
\caption{Theoretical and Simulation Results for Population Size $n=1000$}
\label{exper1}
\end{figure}

\begin{figure}[h]
\centering
\includegraphics[width=2.5 in]{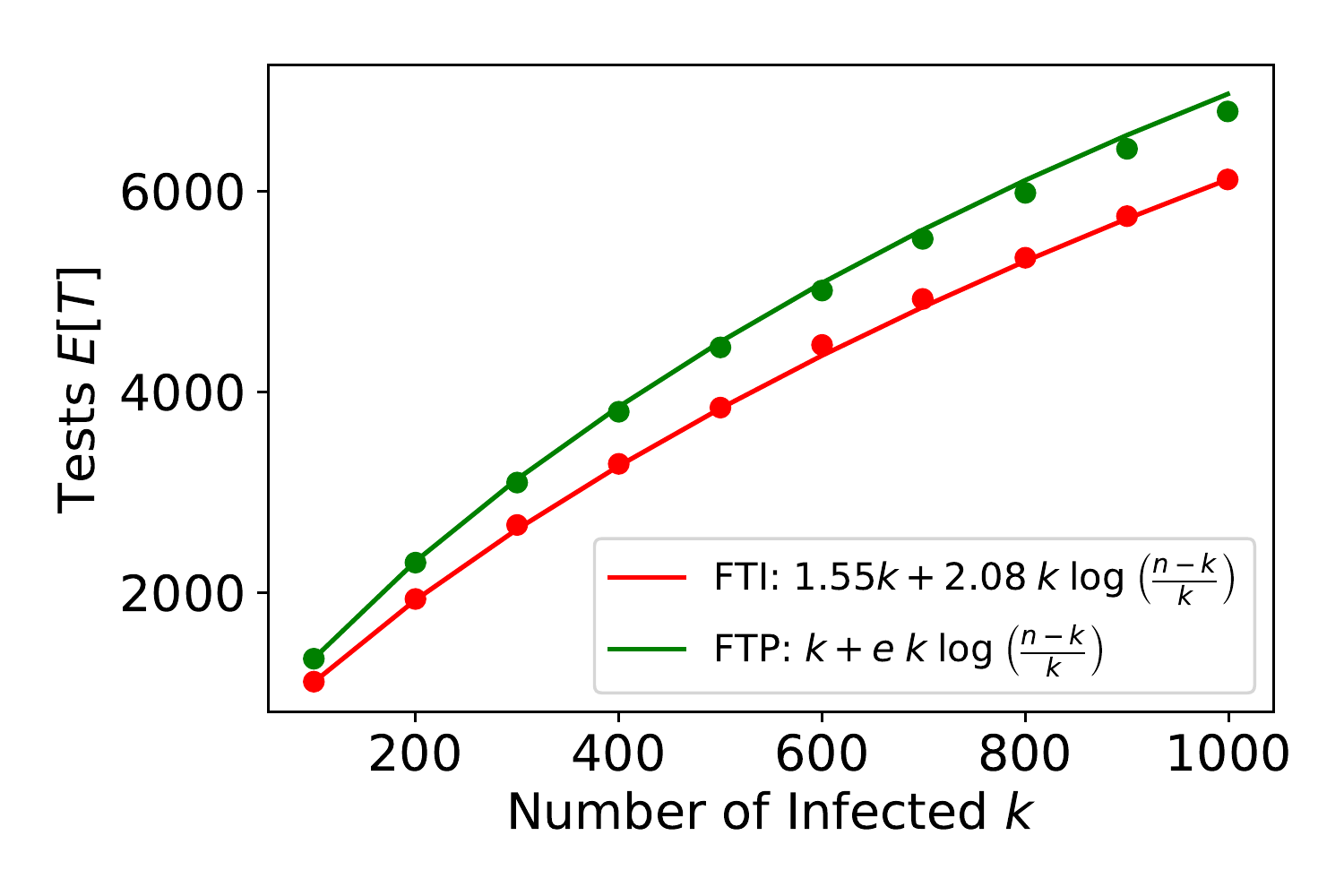}
\vspace{.2mm}
\caption{Theoretical and Simulation Results for Population Size $n=10000$}
\label{exper2}
\end{figure}

In each simulation, the population size $n$, number of infected individuals $k$ is fixed and assignment of individuals to tests are fixed. The 
set of infected individuals is varied by picking $k$ out of $n$ at random. The value of $k$ is varied from $10-100$ in steps of $10$ when the population $n=1000$ and from $100-1000$ in steps of $100$ when the population $n=10000$. The expected number of tests needed to find the infected individuals is determined. This process is repeated $1000$ times. The expected number of testes needed is computed for each run. Each data point shows the expected value and the solid line shows the theoretical value derived. Figure \ref{exper1} shows the result for a population of $n=1000$. There is excellent agreement between the theoretical and simulation results for both the randomization strategies. Figure \ref{exper2} shows the same result for a population of $n=10000.$

\subsection{Experiments with the Binomial Infection Model}
\begin{figure}[h]
\centering
\includegraphics[width=2.5 in]{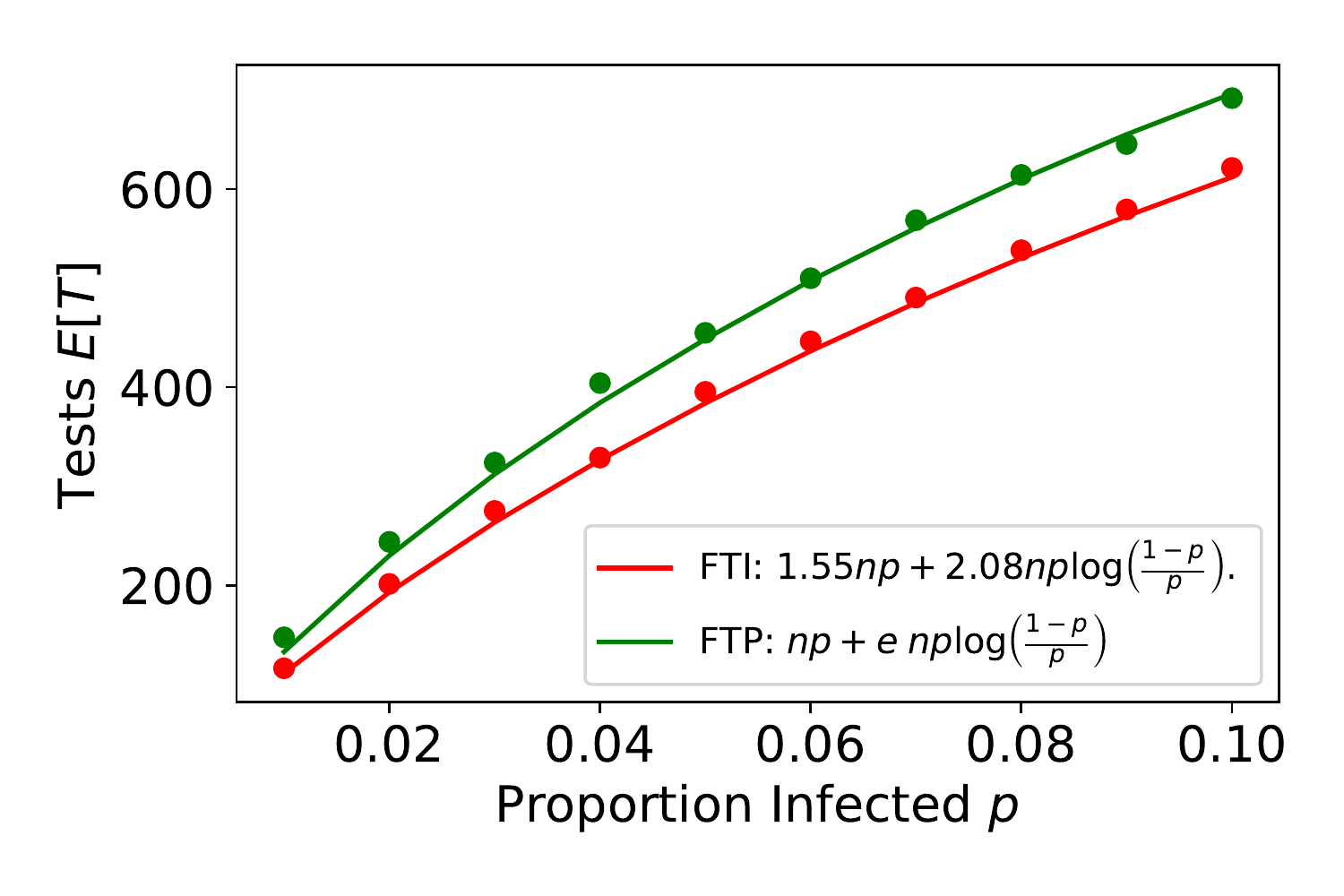}
\vspace{.2mm}
\caption{Theoretical and Simulation Results for Population Size $n=1000$ for the Binomial Model}
\label{exper3}
\end{figure}

\begin{figure}[h]
\centering
\includegraphics[width=2.5 in]{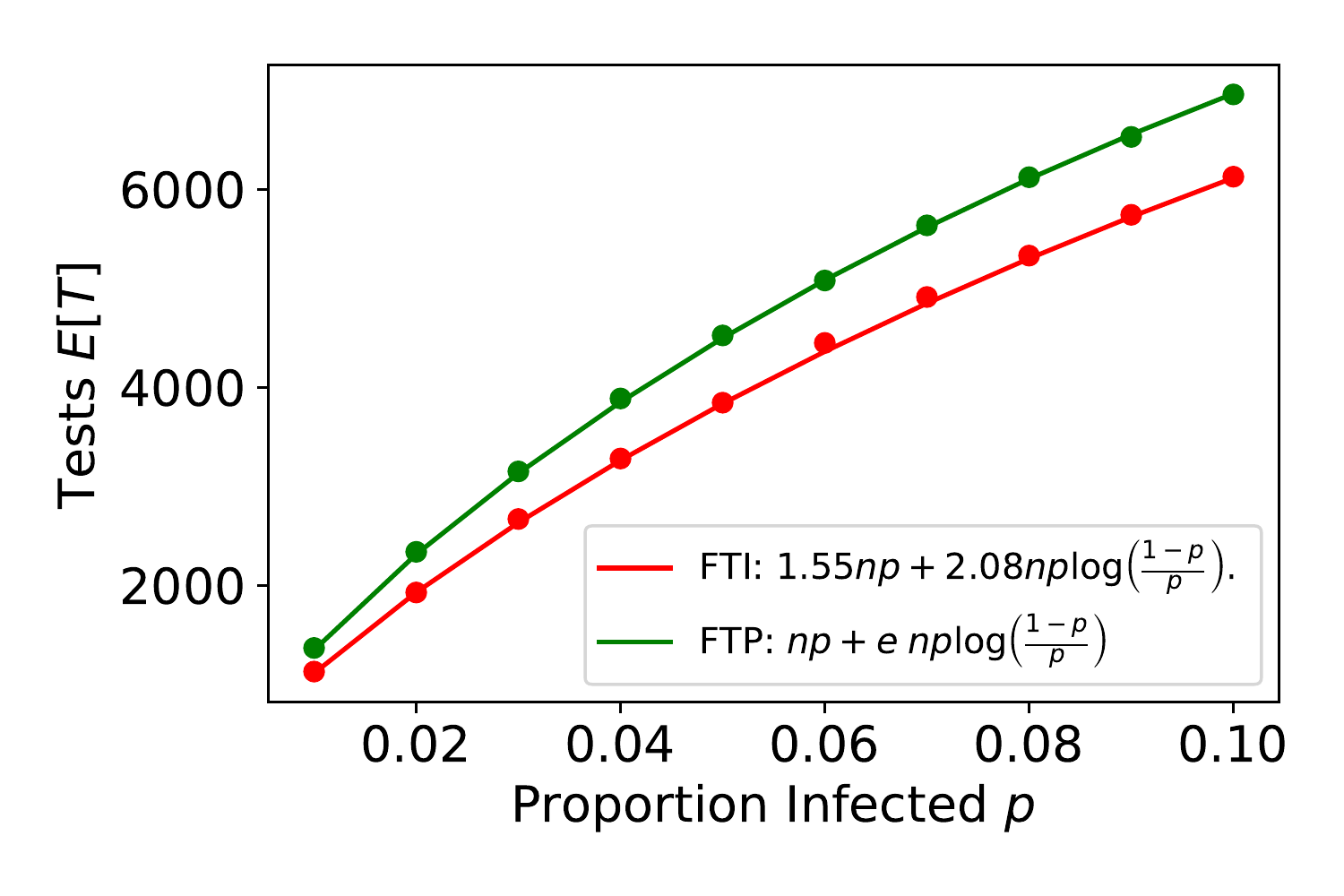}
\vspace{.2mm}
\caption{Theoretical and Simulation Results for Population Size $n=10000$ for the Binomial Model}
\label{exper4}
\end{figure}

We repeat the same experiments but instead of fixed $k$, we assume the binomial infection model where each of the $n$ individuals is infected with probability $p$. We experiment with $n=1000$ (Figure \ref{exper3}) and $n=10000$ (Figure \ref{exper4}) and vary $p$. {\em When designing the pool we only know $n$ and $p$ and not the actual number of infected individuals in a test run}. The two stage tests are designed as outlined in Section \ref{binmodel} by using ${\overline k} = np$ instead of $k$. We vary $p$ from $1\%$ to $10\%$ in steps of $1\%$ for both population sizes. In each run the actual number of infected individuals varies. We again average $1000$
experiments and plot the mean along with theoretically computed curves from Section \ref{binmodel}. Note the agreement between the simulation results and the theoretical expected number of tests in all cases.

\section{Conclusion}
We consider three different randomization schemes for two stage testing and using very simple analysis, show that
Fixed Number of Tests per Individual (FTI) outperforms other randomization schemes.
We show that small differences in the randomization process leads to different performance
results even though asymptotically all the tests are $O\left (k \log \left( \frac{n}{k} \right) \right)$.
We are currently working on extending our model and analysis to testing models where 
tests are not perfect as well as the case 
the accuracy of the result of testing a pooled sample is a function of the number of individuals in the pool. 
\printbibliography
\end{document}